\documentclass[11pt,a4paper]{article}
\usepackage{amsfonts}
\usepackage{graphicx}
\usepackage{amsmath}
\usepackage{hyperref}
\usepackage{enumerate}
\usepackage{amsmath,amssymb}
\usepackage{slashed,mathrsfs}
\usepackage{feynmp,subfigure}
\usepackage{verbatim,graphicx}
\usepackage[sub,ovp]{psfragx}
\usepackage{overpic}

\RequirePackage{mathrsfs} \RequirePackage[sc]{mathpazo}
\RequirePackage{wasysym} \RequirePackage{setspace}
\newcommand{\be}{\begin{equation}}
\newcommand{\ee}{\end{equation}}
\newcommand{\bea}{\begin{eqnarray}}
\newcommand{\eea}{\end{eqnarray}}
\input{tcilatex}

\begin{document}

\title{ \rightline{\mbox{\small
{LPHE-MS-11-03/CPM-11-03}}}\textbf{On Fermion Mass Hirerachies in MSSM-like
Quiver Models with Stringy Corrections}{\ }}
\author{A. Belhaj$^{1,2,3}$\thanks{%
belhaj@unizar.es}, M. Benhamza$^{1}$\thanks{%
benhamzam@gmail.com}, S.E. Ennadifi$^{1}$\thanks{%
ennadifis@gmail.com}, S. Nassiri$^{1}$\thanks{%
nassiris@gmail.com}, E.H. Saidi$^{1,2,3}$\thanks{%
h-saidi@fsr.ac.ma} \\
%EndAName
\\
{\small $^{1}$Laboratory of High Energy Physics, Modeling and Simulation,
Faculty of Science, Rabat, Morocco} \\
{\small $^{2}$Centre of Physics and Mathematics, CPM-CNESTEN, Rabat, Morocco 
}\\
{\small $^{3}$Groupement National de Physique des Hautes Energies, Si\`{e}ge
focal: FSR, Rabat, Morocco } }
\maketitle

\begin{abstract}
Using instanton effects, we discuss the problem of fermion mass hierarchies
in an MSSM-like Type IIA orientifolded model with $\mbox{U(3)}\times %
\mbox{Sp(1)}\times \mbox{ U(1)}\times \mbox{U(1)}$ gauge symmetry obtained
from intersecting D6-branes. In the corresponding four-stack quiver, the
different scales of the generated superpotential couplings offer a partial
solution to fermion mass hierarchies. Using the known data with neutrino
masses $m_{v_{\tau }}$ $\lesssim $ $2$ $eV$, we give the magnitudes of the
relevant scales.

\textbf{Keywords}: Type IIA superstring, Instanton effects and Yukawa
couplings.
\end{abstract}

\newpage

\section{Introduction}

It has been recognized that important particle physics ingredients including
gauge symmetry and chiral matter can be realized in Type II superstring
models using the mechanism of intersecting D-branes \cite{1,2}. More
recently, this method has been extensively progressed and it is now possible
to investigate some stringy effects which could give rise to deeper details
seen in particle physics. In particular, D-brane instantons wrapping non
trivial cycles in the internal manifold have been particularly explored in
this regard \cite{3,4}. They give non-perturbative superpotential
corrections which could explain the large mass hierarchies including the
smallness of neutrino masses in the standard model (SM) \cite{5,6,7,8,9,10}.

Many works have been done along these lines using configurations realizing
MSSM-like orientifolded models based on quiver approach. In this way, rather
than considering full string theory models, working at the level of quivers
allow for dealing with many important physical effects. Indeed, one could
examine the possible presence or absence of couplings and other physical
effects by considering quantum numbers associated with the quiver
configuration data. More precisely, many investigations have been performed
for MSSM-like orientifolded models satisfying necessary constraints allowing
for hierarchical mass terms for all three families of quarks and leptons.
This can be approached using either extra singlet fields that string theory
compactifications contain or via Euclidean D-brane wrapping non trivial
cycles producing E-instantons which we are interested in here \cite{11,12}.
They exhibit the so called uncharged modes consisting of fermion modes $%
\theta ^{\alpha },\overline{\theta }^{\overset{.}{\alpha }}$ preserving half
of the bulk supersymmetry breaking $N=2$ to $N=1$ and the four bosonic modes 
$x^{\mu }$, which are associated with breakdown of four-dimensional Poincar%
\'{e} invariance. These kinds of uncharged zero modes appear if the
instantons wrap non-rigid three-cycles in the internal manifold. However,
the charged fermionic zero modes $\lambda _{a},\bar{\lambda _{b}}$ appearing
at the intersection between E-instantons and D-branes are crucial in the
determination of superpotential corrections to four dimensional effective
field theories. The non-perturbative contributions are given by performing
the path integral over all fermionic zero modes. To get superpotential
terms, one must ensure that all uncharged fermionic zero modes, apart from $%
x^{\mu }$ and $\theta ^{\alpha }$, are projected out or lifted. In type IIA
superstring for instance, this can be done by considering E2-instantons
obtained from D2-branes wrapping rigid orientifold-invariant 3-cycles
embedded in the internal space \cite{3,4}. These are referred to as rigid $%
O(1)$ instantons, which will be discussed here. In this picture, when
instanton effects do not give rise to phenomenological inconveniences, they
therefore represent a way in which one can give rise to viable mass
hierarchies for the quarks and leptons as well as the smallness of neutrino
masses.

In this paper we contribute to these activities by performing a discussion
of mass hierarchies based on a Type IIA four-stack quiver orientifolded
model in the presence of E2-instanton effects. More precisely, we focus on $%
\mbox{U(3)}_{a}\times \mbox{Sp(1)}_{b}\times \mbox{U(1)}_{c}\times %
\mbox{U(1)}_{d}$ quiver gauge theory based on intersecting D6-brane
configurations with instanton corrections to the corresponding effective
superpotential. For simplicity, we consider $O(1)$-instantons wrapping rigid
oriontifold-invariant cycles and carrying the right global $\mbox{U(1)}%
_{c,d} $ charges through the $\lambda _{c,d},\bar{\lambda}_{c,d}$. Then we
give the corresponding MSSM-like quiver with the instanton corrections.
Compared to perturbatively allowed superpotential terms of the heavy fields,
the scales of the induced terms of the light fields depend on the
suppression factors of E2-instantons which may induce them with some
combinations with the string scale $M_{s}$ through higher order terms. This
mechanism offers a stringy framework to explain the fermion mass hierarchies
in the quiver method. Including the left-handed neutrino masses in the
analysis, we get the order of magnitude of the relevant instanton effects
and recover the string scale upper bound $M_{s}\lesssim 10^{14}GeV$.

The organization of this paper is as follows. In section 2, we build a
quiver gauge theory realizing MSSM-like orientifolded models based on four
stacks of intersecting D6-branes. In section 3, we study the fermion mass
hierarchies by introducing the possible stringy corrections to the
corresponding superpotential using O(1) E2-instantons then we give the
corresponding quiver. Their suppression magnitudes have been estimated.
Section 4 will be devoted to our discussions.

\section{Quiver model}

In this section, we will give a quiver model based on four-stack of
D6-branes embedded in IIA superstring moving on orientifolded geometry. The
gauge theory lives on D6-branes wrapping four-dimensional Minkowski space
and non trivial 3-cycles in the internal manifold \cite{13,14}. It is
recalled that D6-branes carry Ramond-Ramond charges that should be cancelled
by the introduction of orientifold geometries related to fixed point loci of
an antiholomorphic involution acting on the internal space. In such quiver
models, the bifundamental chiral matter arises at the non-trivial
intersection of two generic D6-branes. It turns out that, one can have
symmetric or anti-symmetric tensor representations where a D6-brane
intersects its image brane under the orientifold action. Furthermore, two
given D6-branes might intersect in multiple points on the compact internal
space, giving rise to multiple families, where the number of families is the
topological intersection number of two 3-cycles belonging to middle
dimensional cohomology. Indeed, a stack of $N$ D6-branes wrapping 3-cycles
gives rise to $\mbox{U}(N)$ gauge symmetry. However, $N$ D6-branes wrapped
on cycles which are homologically-fixed or pointwise-fixed by the
orientifold action give rise to $\mbox{Sp}(N)$ and $\mbox{SO}(N)$ gauge
symmetry, respectively. Since $\mbox{Sp}(1)$ is isomorphic to $\mbox{SU}(2)$%
, the $\mbox{SU(2)}_{L}$ of the MSSM-like models can be realized as a $%
\mbox{Sp}(1)$ arising from a D6-brane stack wrapping an orientifold
invariant 3-cycle. In this regard, the SM gauge symmetry and the matter
content can be accommodated in the gauge group $\mbox{U}(3)\times \mbox{Sp}%
(1)\times \mbox{U}(1$) obtained from three stacks of D6-branes. Enlarging
the gauge group can make the model richer. This can be done by assuming that
one has an appropriate middle dimensional cohomology generated by non
trivial 3-cycles. The model we consider here relies on a four stacks of
D6-branes giving rise to the following gauge symmetry 
\begin{equation}
\mbox{U(3)}_{a}\times \mbox{Sp(1)}_{b}\times \mbox{U(1)}_{c}\times %
\mbox{U(1)}_{d}  \label{eq1}
\end{equation}%
with a gauged flavor group $\mbox{U(1)}_{d}$ distinguishing various quarks
from each others. Roughly, the tadpole conditions imply the vanishing of
non-abelian anomalies, while abelian and mixed anomalies are canceled via
the Green-Schwarz mechanism. Generically, the anomalous $\mbox{U(1)}%
^{^{\prime }}s$ acquire a mass and survive only as global symmetries, which
forbid various couplings on the perturbative level. Since the SM gauge
symmetries contain the abelian symmetry $\mbox{U(1)}_{Y}$ , and the $%
\mbox{Sp}(1)$ does not exhibit a $\mbox{U(1)}_{b}$ which could contribute to
the hypercharge, one requires that a linear combination of $\mbox{U(1)}_{Y}=%
\underset{k}{\overset{a,c,d}{\sum }}q_{k}\mbox{U(1)}_{k}$ remains massless.
Thus, the resulting gauge group in four-dimensional spacetime can be written
as 
\begin{equation}
\mbox{SU(3)}_{a}\times \mbox{Sp(1)}_{b}\times \mbox{U(1)}_{Y}.  \label{eq2}
\end{equation}%
Given the above gauge symmetry based on D6-brane configurations, we can
construct a quiver describing MSSM-like orientifolded model. Indeed,
vanishing of anomalies which we require to be satisfied are used to fit the
matter content involving two up quarks, one down quark charged under the $%
\mbox{U(1)}_{c\text{ }}$ gauge symmetry and the opposite arrangement charged
under $\mbox{U(1)}_{d}.$ The model we present here relies on a particular
intersection numbers of 3-cycles in middle dimensional cohomology. For that,
we choose the following intersections\footnote{%
We have not included those involving $b^{\ast }=b$} 
\begin{eqnarray}
\text{ }I_{D_{a}D_{b}} &=&3,\text{ \ \ }I_{D_{a}D_{c}}=-2,\text{ \ }%
I_{D_{a}D_{c^{\ast }}}=-1,  \notag  \label{eq3} \\
\text{ }I_{D_{a}D_{d}} &=&-1,\text{ \ \ }I_{D_{a}D_{d^{\ast }}}=-2,\text{ }%
I_{D_{d}D_{b}}=3, \\
I_{D_{d}D_{c^{\ast }}} &=&-3.\text{ \ }  \notag
\end{eqnarray}%
The other intersection numbers are set to zero. The chiral spectrum and the
gauge symmetry can be represented in the table 1 together with their
identification with SM matter fields. 
\begin{table}[th]
\begin{center}
\begin{tabular}{|c|c|c|c|c|c|c|c|c|}
\hline
$Sector$ & $ab$ & $ac$ & $ac^{\ast }$ & $ad$ & $ad^{\ast }$ & $db$ & $%
dc^{\ast }$ & $bc,bc^{\ast }$ \\ \hline
$Fields$ & $q^{i}$ & $\overline{u}^{2,3}$ & $\overline{d}^{3}$ & $\overline{u%
}^{1}$ & $\overline{d}^{1,2}$ & $l^{i}$ & $\overline{e}^{i}$ & $H_{d},$ $%
H_{u}$ \\ \hline
$Rep$ & $3(3,\overline{2})$ & $2(\overline{3},1)$ & $1(\overline{3},1)$ & $1(%
\overline{3},1)$ & $2(\overline{3},1)$ & $3(1,\overline{2})$ & $3(1,1)$ & $%
1(1,2)$ \\ \hline
$Q_{a}$ & $1$ & $-1$ & $-1$ & $-1$ & $-1$ & $0$ & $\ 0$ & $0,$ $0$ \\ \hline
$Q_{c}$ & $0$ & $1$ & $-1$ & $0$ & $0$ & $0$ & $-1$ & $1,-1$ \\ \hline
$Q_{d}$ & $0$ & $0$ & $0$ & $1$ & $-1$ & $1$ & $-1$ & $0,\ 0$ \\ \hline
$Y$ & $1/6$ & $-2/3$ & $1/3$ & $-2/3$ & $1/3$ & $-1/2$ & $1$ & $-1/2,1/2$ \\ 
\hline
\end{tabular}%
\end{center}
\caption{The spectrum and their $U(1)_{a,c,d}$ charges for $Y=\frac{1}{6}%
Q_{a}-\frac{1}{2}Q_{c}-\frac{1}{2}Q_{_{d}}$. The index $i(=1,23)$ denotes
the family index}
\end{table}
In this description, all leptons and the two Higgs doublets are charged
under $\mbox{U(1)}_{c,d}$ symmetries while for quarks only their right
handed partners are. The model with four stacks can be encoded in a quiver
where each node represents a D6-brane and the links between them indicate
their chiral intersections. The quiver summarizing the above spectrum with
the two Higgses is shown in figure 1

\begin{figure}[tbph]
\begin{center}
\includegraphics[width=9cm]{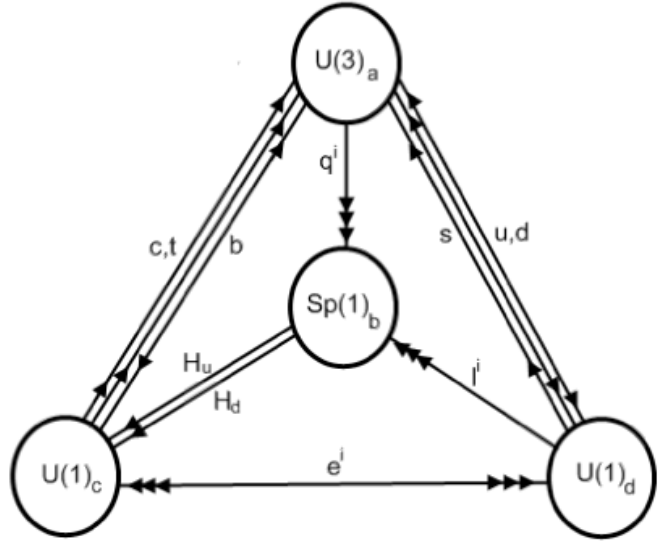}
\end{center}
\caption{Quiver for intersecting D6-branes: Circles denote D6-branes while
bold lines indicate the chiral spectrum. Arrows directions indicate
fundamental (antifundamental) representations of U($N$) gauge group. }
\end{figure}
Since any realistic string vacua has to exhibit the phenomenologically
desired terms in four dimensional effective superpotential, we require the
presence of all the MSSM Yukawa couplings and the absence of the
phenomenologically undesired couplings terms on perturbative level.

In what follows, the $\mbox{U(1)}_{c,d}$ symmetries will be used to select
the candidate terms for the superpotential. Taking into account the charges
presented in the table 1, we can write down the all allowed interaction
terms generating the possible Yukawa couplings. It turns out that, the only
fields involved in the allowed perturbative interaction terms are those
charged under the $\mbox{U(1)}_{c}$ gauge symmetry. We interpret them as the
the heavy quarks $c,t,b$, leptons $e^{i}$ and the $\mu $ term for $H_{u},$ $%
H_{d}$. This could be seen at the level of their associated quantum numbers 
\begin{eqnarray}
&&H_{u\left( 0,-1,0\right) }q_{\left( 1,0,0\right) }c_{\left( -1,1,0\right)
},\text{ }H_{u\left( 0,-1,0\right) }q_{\left( 1,0,0\right) }t_{\left(
-1,1,0\right) },\text{ }H_{d\left( 0,1,0\right) }q_{\left( 1,0,0\right)
}b_{\left( -1,-1,0\right) },\text{ }  \notag  \label{eq4} \\
&&H_{d\left( 0,1,0\right) }l_{\left( 0,0,1\right) }^{i}e_{\left(
0,-1,-1\right) }^{i},\text{ } \\
&&H_{u\left( 0,-1,0\right) }H_{d\left( 0,1,0\right) }.  \notag
\end{eqnarray}%
Here the index $i$ denotes the family index and the subscript indicates the
charge under the $\mbox{U(1)}_{a,c,d}$ symmetries. Our quiver naturally
allows for the following couplings 
\begin{equation}
y_{c}H_{u}q\overline{c}+y_{t}H_{u}q\overline{t}+y_{b}H_{d}q\overline{b}%
+y_{e^{i}}H_{d}l^{i}\overline{e}^{i}+\mu H_{u}H_{d}  \label{eq5}
\end{equation}%
where $y_{c,t,b}$ are coupling constants accounting for hierarchies between
these terms. The absent couplings, which are phenomenologically desired,
corresponding to fermions charged under $\mbox{U(1)}_{d}$ violate the $%
\mbox{U(1)}_{c,d}$ symmetries. These will be refered to\ the light fermions $%
u,d,s,v$ that will be yielded massless, unless some kind of new effects in
the defined effective field theory break the remnants abelian symmetries $%
\mbox{U(1)}_{c,d}$. In string theory, the natural candidate non-perturbative
effects to violate these $\mbox{U(1)}_{c,d}$ symmetries are instantons
arising from euclidean D-branes coupling to these fields \cite{15,16,17,18}.
They can potentially destabilize the vacuum or lead to new effects in the
four-dimensional effective action. In what follows, we will discuss the
implication of such non perturbative effects in our quiver model. In
particular, we will consider orientifolded invariant 3-cycles on which
D2-branes can wrap to make rigid $O(1)$ instantons. Then, we discuss fermion
mass hierarchies in such instanton configurations.

\section{ Stringy corrections and mass hierarchies}

\subsection{Instanton corrections}

In this section we introduce the effects of spacetime instantons on the
above described model. In type IIA superstring model, such non-perturbative
effects are generated in principle from D0, D2 and D4-branes wrapping one,
three and five-cycles respectively. However, since a Calabi-Yau manifold
does not have any continuous one and five-cycles, the only relevant
instantons are D2-branes wrapped on three-cycles in the internal manifold 
\cite{3,4}. Roughly, the resulting D2-instanton action takes the follwing
form 
\begin{equation}
S_{E}=S(\Lambda ,\Phi _{n})+S_{E}^{cl}.  \label{eq6}
\end{equation}%
In this equation, $S(\Lambda ,\Phi _{n})$ describes all terms involving
fermionic instanton zero modes and four dimensional charged matter fields $%
\Phi _{n}$ while $S_{E}^{cl}$ is the Dirac Born-Infeld action on D2-brane
wrapping a 3-cycle $\Sigma $ in the presence of the WZ term. This action
reads as 
\begin{equation}
S_{E}^{cl}\sim V(\Sigma )+i\int_{\Sigma }C_{3}  \label{eq7}
\end{equation}%
where $C_{3}$ is the R-R 3-from coupled to the D2-brane. In four dimensions,
the instanton contributions to the low energy effective theory can be
obtained by performing the Grassmann path integral over all fermionic zero
modes $\Lambda $: 
\begin{equation}
S_{E}^{4d}\left( \Phi _{n}\right) =\int \left[ D\Lambda \right] e^{-S_{E}}={%
\underset{n}{\Pi }}\Phi _{n}e^{-S_{E}^{cl}}.  \label{eq8}
\end{equation}%
The classical instanton action $e^{-S_{E}^{cl}}$ can absorb the $\mbox{U(1)}$%
's charge excess of the matter fields operator ${\underset{n}{\Pi }}\Phi
_{n} $ \cite{15,16,17}. Under such abelian symmetries the transformation
property of the exponential instanton action is 
\begin{equation}
\text{ }e^{-S_{E}^{cl}}\longrightarrow e^{iQ(E2)\Lambda }e^{-S_{E}^{cl}}
\label{eq9}
\end{equation}%
where $Q(E2)$ represents the amount of the U(1)-charge violation by the $E2$%
-instantons. Their microscopic origin is in the extra fermionic zero modes $%
\Lambda $ living in the intersection of the E2-branes with the $D6$ branes.

Instead of being general, let us consider a concrete configuration with $E2$%
-intantons wrap rigid orientifold-invariant 3-cycles. In this case, the $%
E2-D6$ and $E2-D6^{\ast }$ are identified. So, an $E2$-instanton
intersecting $D6_{c,d}$-branes can induce the desired couplings of the
matter fields operator ${\underset{n}{\Pi }}\Phi _{n}=H_{u}qu,$ $H_{d}qd,$ $%
H_{d}qs,\ldots $. The $\mbox{U(1)}_{c,d}$ charges which are carried by the
instanton action have their origin in the intersection $E2-D6_{c,d}$
pattern. For an instanton wrapping a 3-cycle with an appropriate number of
times, this can be made exactly opposite to the total charge of the operator 
${\underset{n}{\Pi }}\Phi _{n}$, so the coupling (8) is also $\mbox{U(1)}%
_{c,d}$-invariant. Examining the $Q_{c,d}$ charge-excess for each term, we
determine the intersection $E2-D6_{c,d}$ pattern giving rise to the right
charged fermionic zero modes $\lambda _{c,d},$ $\overline{\lambda }_{c,d}$
compensating the $Q_{c,d}$ charge-excess of the desired couplings. Indeed,
the up-quark coupling term 
\begin{equation}
H_{u\left( 0,-1,0\right) }q_{\left( 1,0,0\right) }u_{\left( -1,0,1\right) }
\label{eq10}
\end{equation}%
violate the $\mbox{U(1)}_{c,d}$ symmetries by one unit as $%
Q_{c}(Hqu)=-1,Q_{c}(Hqu)=1.$ Then is generated by an instanton $E2_{u}$
intersecting $D6_{c,d}$ branes with the following intersection numbers 
\begin{equation}
\text{ }I_{E_{u}D_{c}}=-1,\text{ \ \ }I_{E_{u}D_{d}}=1  \label{eq11}
\end{equation}%
which gives rise to two charged modes $\lambda _{c}^{u\text{ }},$ $\overline{%
\lambda }_{d}^{u}:$ 
\begin{equation}
\text{ }Q_{c}(\lambda _{c}^{u\text{ }})=-N_{c}I_{E_{u}Dc}=1,\text{ \ }Q_{d}(%
\overline{\lambda }_{d}^{u})=-N_{d}I_{E_{u}D_{d}}=-1.\text{ }  \label{eq12}
\end{equation}%
These charges compensate the charge-excess of the up-quark term. In this
case, the corresponding E2-instanton action takes the following form 
\begin{equation}
\text{ }S_{E_{u}}=S_{E_{u}}^{cl}+\lambda _{c}y_{u}H_{u}qu\overline{\lambda }%
_{d}.  \label{eq13}
\end{equation}%
Performing the integration over the all fermionic zero mode, one get the
desired term 
\begin{equation}
\int d^{4}xd^{2}\theta d\lambda _{c}d\overline{\lambda }%
_{d}e^{-S_{E_{u}}^{cl}+y_{u}\lambda _{c}H_{u}qu\overline{\lambda }%
_{d}}=e^{-S_{E_{u}}^{cl}}\int d^{4}xd^{2}\theta d\lambda _{c}d\overline{%
\lambda }_{d}y_{u}\lambda _{c}H_{u}qu\overline{\lambda }%
_{d}=e^{-S_{E_{u}}^{cl}}y_{u}H_{u}qu.  \label{eq14}
\end{equation}%
Analogous analysis can be done for the down and strange quark coupling terms 
\begin{equation}
H_{d\left( 0,1,0\right) }q_{\left( 1,0,0\right) }d_{\left( -1,0,-1\right) },%
\text{ }H_{d\left( 0,1,0\right) }q_{\left( 1,0,0\right) }s_{\left(
-1,0,-1\right) }.  \label{eq15}
\end{equation}%
This can be generated by an $E2_{d,s}$-instanton intersecting the $D6_{c,d}$%
-branes with the following intersection numbers 
\begin{equation}
\text{ }I_{E_{d,s}D_{c}}=1,\text{ \ \ }I_{E_{d,s}D_{d}}=-1  \label{eq16}
\end{equation}%
and producing the two charged modes $\overline{\lambda }_{c}^{d,s\text{ }},$ 
$\lambda _{d}^{d,s}$. For the sector of neutrinos, the following higher
order term 
\begin{equation}
H_{u\left( 0,-1,0\right) }l_{\left( 0,0,1\right) }^{i}H_{u\left(
0,-1,0\right) }l_{\left( 0,0,1\right) }^{i}  \label{eq17}
\end{equation}%
could be generated by a $E2_{v}$-instanton intersecting $D6_{c,d}$-branes.
The corresponding intersection numbers are given by 
\begin{equation}
\text{ }I_{E_{v}D_{c}}=-2,\text{ \ \ }I_{E_{v}D_{d}}=2.\text{ }  \label{eq18}
\end{equation}%
This leads to the charged modes $\lambda _{c}^{v\text{ }},$ $\bar{\lambda}%
_{d}^{v}$. Integrating over the all above fermionic zero modes as in (\ref%
{eq14}), we get the missing superpotential terms at the four-dimensional
theory 
\begin{equation}
y_{u}e^{-S_{E_{u}}^{cl}}H_{u}q\overline{u}+e^{-S_{E_{d,s}}^{cl}}(y_{d}H_{d}q%
\overline{d}+y_{s}H_{d}q\overline{s}%
)+y_{v^{i}}e^{-S_{E_{v^{i}}}^{cl}}M_{s}^{-1}\left( H_{u}l^{i}\right) ^{2}
\label{eq19}
\end{equation}%
where the coupling constants $y_{d,s}$ can account for hierarchies between
the $d$ and $s$ quarks having the same instanton suppression. The neutrinos
superpotential term is highly suppressed by $1/M_{s}$ factor and once Higgs
fields get a Vev $v_{u}$ this operator gives rise directly to Majorana
masses for the left-handed neutrinos depending on the scale $M_{s}$ taken as
the low string scale at which neutrino masses have origin.

The quiver illustrating these instanton intersection patterns with their
appropriate charged fermionic zero modes is depicted in the figure2. 
\begin{figure}[tbph]
\begin{center}
\includegraphics[width=9cm]{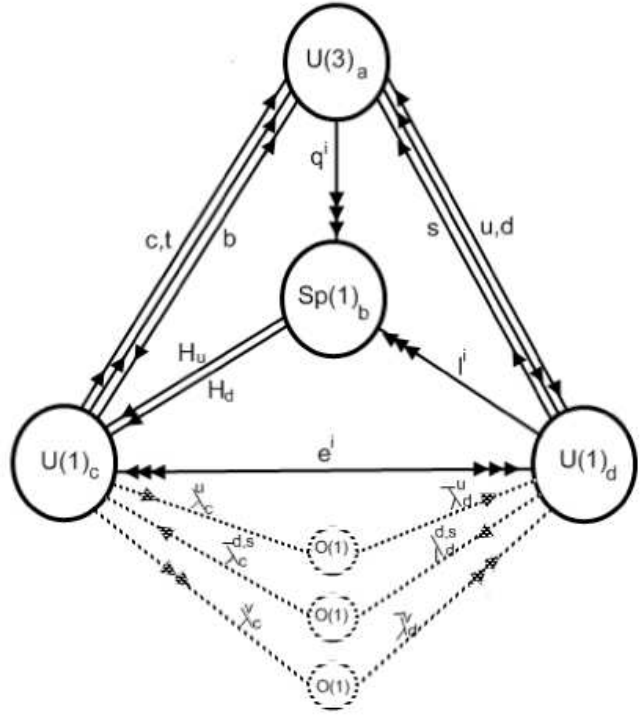}
\end{center}
\caption{Quiver for intersecting D6-branes and E2 -instantons: Dotted
circles denote the $O(1)$-instantons and dotted lines indicate their
intersections with the $D6_{c,d}$-branes.}
\end{figure}

\subsection{Instanton correction magnitudes}

Recall that the exponential suppression effects can be derived from the
classical instanton actions. Their magnitudes depend on the internal
geometry on which the quiver model is based. In particular, they depend on
the complex structure moduli space described by the volume of the 3-cycles
wrapped by the relevant instantons. To have an idea about the induced mass
hierarchies, we need to estimate the different instanton factors. Although
such approach forces to consider deeper details of the fully defined string
model, we can get an order of magnitude of the different suppression factors
by refereing to the known data. Effectively, after the Higgs fields break
the electroweak symmetry at the usual scale $v_{u}^{2}+$\ $v_{d}^{2}=\left(
246GeV\right) ^{2}$, \ we take the quark couplings terms of eq(\ref{eq19})
together with the quark coupling terms of eq (\ref{eq5}) with some
assumption on their scalar-fermion couplings to derive the expected
magnitudes. For that, using some combination of the quark masses where their
net scalar-fermion couplings effect could be neglegted, one allows for
getting approximate values of the corresponding suppression factors. For the 
$u$-quark suppression factor, assuming $y_{u}\sim y_{d}$ and $y_{s}\sim
y_{c} $, we find 
\begin{equation}
e^{-S_{E_{u}}^{cl}}\simeq \frac{m_{u}m_{s}}{m_{d}m_{c}}\sim 3.10^{-2}
\label{eq20}
\end{equation}%
While for the $d$ and $s$-quark suppression factors with $y_{d}+y_{s}\sim
y_{b}$, we get 
\begin{equation}
\text{ }e^{-S_{E_{d,s}}^{cl}}\simeq \frac{m_{d}+m_{s}}{m_{b}}\sim 2.10^{-2}%
\text{ }.\text{ }  \label{eq21}
\end{equation}%
Similarly, using the known data with neutrino masses upper bound $m_{v\tau }$
$\leqslant 2eV$ we find the neutrino high suppression factor%
\begin{equation}
\frac{e^{-S_{E_{v^{i}}}^{cl}}}{M_{s}}\sim 10^{-14}GeV^{-1}.  \label{eq21}
\end{equation}%
Considering the usual value of the string scale, namely $M_{s}\simeq
10^{18}GeV$, the E2-instanton induced operator only leads to subleading
corrections to the neutrino masses. Their observed order can be obtained by
lowering the string scale due to large internal dimensions \cite{19,20}.
Lowering the string scale down to the $TeV$ scale, this leads to the
interesting features at the LHC. From (\ref{eq21}), we can deduce the string
scale upper bound 
\begin{equation}
M_{s}\leqslant 10^{14}\ GeV.
\end{equation}%
Investigating the above instanton suppression factors (\ref{eq20}) and (\ref%
{eq21}) and specializing the case that the D6-branes and the E2-branes wrap
factorizable three-cycles of toroidal orientifolds, the volume of these
instantons could give an insight of the detailed description of geometric
background of the internal manifold. We believe that this connection
deserves to be studied further.

\section{ Conclusions and Discussions}

In this paper, we have discussed the fermion mass hierarchies in a local
four-stack intersecting MSSM-like D-brane quiver models using non
perturbative stringy effects. More precisely, we have focused on a
four-stack of D6-brane configurations in type IIA orientifolded geometries
giving rise to the MSSM-like spectrum without right handed-neutrinos
including rigid O(1) E2-instantons. In particular, we have given a quiver
model from which we have shown that some perturbatively absent coupling
terms can be generated from D2-branes wrapping 3-cycles belonging to middle
dimensional cohomology of the internal space.

Analyzing the quiver allows for the determination of perturbatively and
non-perturbatively contributions to the superpotential using the abelian
symmetries obeyed at the perturbative regime. In this approach, attributing
the perturbatively realized terms to the heavy fermions and the remaining
missing desired terms to the light fermions reflects interesting
hierarchies. The latters are induced and exponentially suppressed by
E2-instantons carrying the right charged fermionic zero modes appearing at
the E2-D6 intersections.

These stringy corrections do not induce the perturbatively forbidden
dangerous proton decay terms through the dimension 5 operators $%
q_{L}q_{L}q_{L}l$ and $u_{R}u_{R}d_{R}e_{R}$. Such kinds of local
intersecting D-brane models have been discussed extensively in \cite{5,6,7,9}%
. These models, which include phenomenological constraints of fast proton
decay are listed in \cite{7}. However, the presented model does not appear
in that classification since it does not involve right-handed neutrinos in
addition to the lepton number violation that arises through $l_{L}l_{L}e_{R}$
and $l_{L}H_{u}$ terms induced by the $E2_{u}$ instanton required to
generate the missing up-quark mass term. Despite, it should be interesting
to make contact with the models given in \cite{7,9}. We believe that this
connection deserves more study. This will be reported elsewhere.

Using the known data with neutrino masses $m_{v_{\tau }}$ $\lesssim $ $2$ $%
eV $, we have given the magnitudes of the relevant instanton effects and the
string scale upper bound $10^{14}$\ $GeV$ imposed to match with the observed
order of the known data.

In the end of this work, a possible discussion could be done in terms of
middle dimensional cohomology of the internal space describing its complex
structure. The analysis depends on the volume of the 3-cycles on which
D2-branes wrap to make rigid O(1) instantons. It should be very nice to find
a geometrical interpretation in terms of quiver data encoded in the middle
dimensional cohomology. We believe that it will be useful to explore
extended Dynking geometries involving more than bosonic vertices. This seems
to be promising in Type IIB D5-branes set-up in the presence of D-strings
wrapping 2-cycles making instantons.

\paragraph{\protect\bigskip Acknowledgement}

AB would like to thank P. Camara for discussions. We thank also   URAC 09/CNRST.

\end{document}